\documentclass[letterpaper]{article} % DO NOT CHANGE THIS
\usepackage{aaai2026}  % DO NOT CHANGE THIS
\usepackage{times}  % DO NOT CHANGE THIS
\usepackage{helvet}  % DO NOT CHANGE THIS
\usepackage{courier}  % DO NOT CHANGE THIS
\usepackage[hyphens]{url}  % DO NOT CHANGE THIS
\usepackage{graphicx} % DO NOT CHANGE THIS
\usepackage{natbib}  % DO NOT CHANGE THIS AND DO NOT ADD ANY OPTIONS TO IT
\usepackage{caption} % DO NOT CHANGE THIS AND DO NOT ADD ANY OPTIONS TO IT
\usepackage{algorithm}
\usepackage{algorithmic}

\usepackage{amsmath}
\usepackage{xcolor}

\newcommand{\consumerq}{\textsc{ConsumerQ}}

\usepackage{tcolorbox}\usepackage{enumitem}

\usepackage{fvextra}

\DefineVerbatimEnvironment{Prompt}{Verbatim}{
    breaklines=true,
    breakanywhere=true,
    fontsize=\footnotesize,
    frame=single,
    framesep=3mm,
    rulecolor=\color{gray},
    breaksymbolleft={},
    breaksymbolright={},
    xleftmargin=0.5em,
    xrightmargin=0.5em
}

\usepackage{newfloat}
\usepackage{listings}
\DeclareCaptionStyle{ruled}{labelfont=normalfont,labelsep=colon,strut=off} % DO NOT CHANGE THIS
\floatstyle{ruled}
\newfloat{listing}{tb}{lst}{}
\floatname{listing}{Listing}

\usepackage{booktabs}

\nocopyright

\title{\emph{``If I Had to Buy Just ONE: Galaxy S26 Ultra''} \\ Auditing AI-Generated Product Recommendations\footnote{This paper was not sponsored by Samsung.}}
\author{
    Lucas G. Uberti-Bona Marin\equalcontrib\textsuperscript{\rm 1},
    Thales Bertaglia\equalcontrib\textsuperscript{\rm 2},
    Giovanni Astante\textsuperscript{\rm 3},
    Bram Rijsbosch\textsuperscript{\rm 1},
    Gijs van Dijck\textsuperscript{\rm 1},
    Anikó Hannák\textsuperscript{\rm 3},
    Gerasimos Spanakis\textsuperscript{\rm 1},
    Konrad Kollnig\textsuperscript{\rm 1}
}
\affiliations{
    \textsuperscript{\rm 1}Law \& Tech Lab, Maastricht University\\
    \textsuperscript{\rm 2}Utrecht University\\
    \textsuperscript{\rm 3}Social Computing Group, Department of Informatics, University of Zurich\\
    \textbf{Corresponding authors:}lucas.uberti-bonamarin@maastrichtuniversity.nl, t.f.costabertaglia@uu.nl
}

\begin{document}

\maketitle

\begin{abstract}
Consumers increasingly use AI chatbots for advice on what to buy. With companies like OpenAI and Google monetising their AI through advertising, this raises difficult questions about the bias and impartiality of such advice.
In response, we conduct an AI audit of popular chatbots using real commercial-advice queries.
First, we curate a dataset of 2,528 real commercial-advice queries (\consumerq{}).
Then, we evaluate 1,536 responses to product queries from popular AI chatbots: ChatGPT (chatbot and API), Google Gemini (chatbot and API), and Google Search (AI Overviews).
We find that ChatGPT expresses a first-person product preference in 79\% of product-recommending responses, compared with 7\% for Gemini and 2\% for AI Overviews, while the products recommended often change across repeated requests.
Displayed sources vary strongly: for the same query, the ChatGPT and Gemini interfaces share only 5.4\% of domains on average, with no domain in common in 76.7\% of comparisons. APIs provide a different view from their corresponding interfaces, with mean domain overlaps of 12.0\% for ChatGPT and 14.8\% for Gemini, and also differ in the types and layers of source information they expose.
Our findings show that neither isolated responses nor API observations can be assumed to represent the commercial advice consumers encounter. Independent audits of AI-mediated commercial advice should therefore account for repeated responses, consumer-facing conditions, and the source layer being observed.
\end{abstract}

\section{Introduction}
AI chatbots are increasingly becoming important intermediaries between consumers and the products they buy.
Unlike traditional search, they do not merely retrieve information but generate an answer: selecting products, synthesising sources, and sometimes even presenting the result as a personal recommendation.
Someone looking for a new phone, for example, may ask which phone has the best camera. Google Search's AI Overview answers that ``\textit{the Oppo Find X9 Ultra is widely rated as the best overall camera phone}'', while ChatGPT responds: ``\textit{my pick right now is ... iPhone 17 Pro Max}''. Which recommendation a consumer receives may therefore depend not only on the available information, but also on how AI companies design their chatbots and assistants.

Such queries are becoming increasingly common. In the largest published measurement of ChatGPT usage from 2025, product and service recommendations already accounted for roughly 2\% of conversations  \cite{chatterji2025how}.
This, too, increasingly creates several potential risks for consumers.
Prior work suggests that conversational and anthropomorphic design of AI assistants can positively affect trust and purchase intentions \cite{konya-baumbachSomeoneOutThere2023, caoAnthropomorphismVirtualInfluencers2026}, while users may over-rely on LLM-generated advice even when it is wrong \cite{spathariotiComparingTraditionalLLMbased2023}.
Generated recommendations are usually not even deterministic: repeated requests can return different products or sources.

At the same time, the neutrality of popular AI assistants is unclear, given potentially conflicting commercial interests in providing certain information over others. Only in August, OpenAI expanded advertising in ChatGPT to 31 European markets, which are most likely to appear for product recommendation queries (in between 10 and 14\% of sessions) \cite{lurieBeginningChatGPTAds2026}.
Popular AI assistants, including ChatGPT, Claude, Grok, and Perplexity, have been found to share consumer data directly with companies like Meta, Google, and TikTok for advertising and analytics purposes \cite{jazlanTrackingConversationsMeasuring2026,girishPrivacyControlsTransparency}.
Moreover, the sources AI assistants can draw on depend on the licensing deals they make with publishers.
For example, both OpenAI and Google have agreements providing real-time access to Reddit data, while OpenAI's partnership with Axel Springer makes content from publications including Politico and Business Insider available for use in ChatGPT.

These concerns have recently gained particular regulatory urgency in the European Union. On 31 August 2026, the European Commission designated ChatGPT as a Very Large Online Search Engine (VLOSE) under the Digital Services Act (DSA) \cite{ec2026designation}. Consequently, ChatGPT is subject to stringent obligations on systemic risks -- including risks to consumer protection in online purchasing -- and to independent audits and researchers' data access \cite{RegulationEU20222022a}.
At the same time, consumer protection authorities in the US and EU have been very active in enforcing against deceptive or misleading advertising, including under the EU Unfair Commercial Practices Directive (UCPD) and the Federal Trade Commission Act.
Understanding what consumers actually encounter, and whether API-based audits capture it accurately, has therefore become relevant not only to researchers but also to the emerging regulatory scrutiny of these systems.

A growing literature has already audited generative search for verifiability and claim fidelity
\cite{liu2023evaluating}, reliance on AI-generated sources \cite{allahamSyntheticSourcesAuditing2026}, divergence from organic ranking and
publisher impact \cite{grossmanHowGenerativeAI2026}, and susceptibility to optimisation and injection \cite{aggarwalGEOGenerativeEngine2024a, ye2026promptinjectionroleconfusion}. Yet, two important gaps remain. First, most works focus on general queries and the properties of cited sources, leaving aspects specific to commercial advice queries understudied. Second, audits are often run through provider APIs using researcher-written template queries. While this may be valid for some analyses, we show that it is not when studying commercial advice. Moreover, consumers do not typically use APIs when seeking product recommendations; therefore, web interfaces should be used to analyse those recommendations, not APIs. This is especially important for generating relevant legal evidence, given that previous research has already highlighted differences between API and user interface responses \cite{schatto-eckrodtChatGPTNewsRecommender2025, kirgisLLMSpiralsDelusion2026a, wangAPIBenchmarkScores2026a}.

We address these two gaps by extracting responses to 117 real product-recommendation queries written by users on the ChatGPT and Gemini consumer interfaces, their corresponding APIs, and Google AI Overviews. This yields 1,755 observations collected from a fixed location, with each interface request paired to an API request.
Our work is guided by three research questions:

\begin{itemize}
    \item RQ1: How do product recommendations vary across popular AI chatbots, including repeated queries?
    \item RQ2: What sources do AI chatbots cite for product recommendations, and how do they vary across providers?
    \item RQ3: To what extent can external auditors (e.g. researchers, NGOs, and regulators) rely on platforms' APIs to audit their consumer AI chatbots?
\end{itemize}

\begin{figure*}
  \centering
  \includegraphics[width=0.9\textwidth]{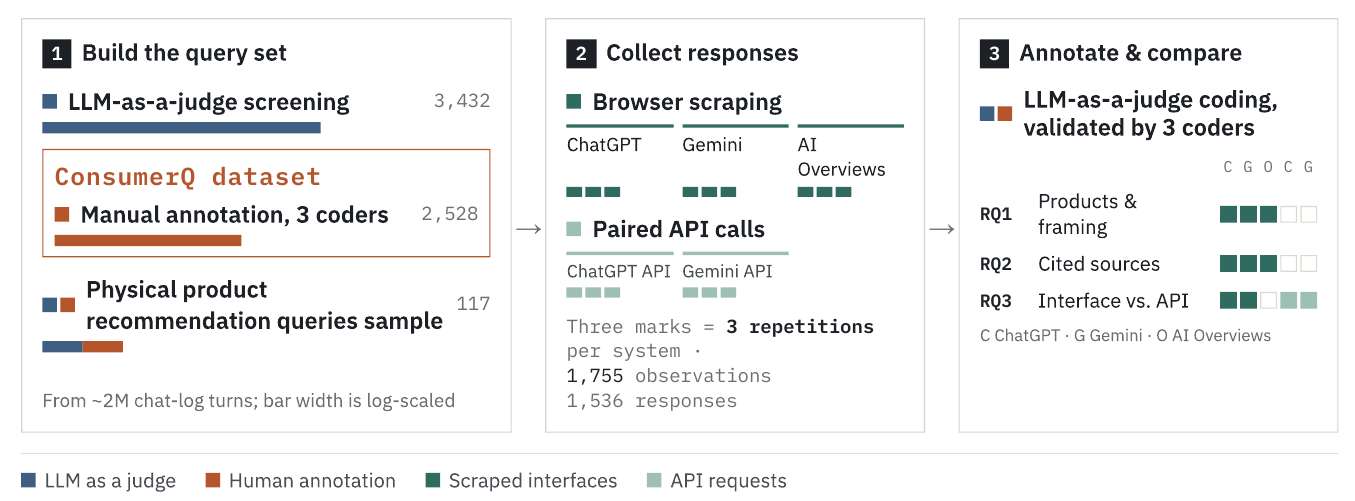}
  \caption{From user chats to product recommendations. We highlight how \consumerq{} is curated, the extracted sample used for analysis and which responses are used to study recommended products and their framing (RQ1), displayed sources (RQ2), and differences between interfaces and API (RQ3)}
  \label{fig:methodology_figure}
\end{figure*}

Our findings follow the paper's three research questions.
\textbf{Providers differ sharply in how confidently they present their recommendations} (RQ1). ChatGPT states a preference in the first person in 79\% of responses, against
7\% for Gemini and 2\% for AI Overviews, and labels some product ``best'' in 73\% of responses against 43\% and 48\%. The same request therefore arrives as a personal recommendation, an undisputed winner, or a list of options, depending on which system was used. This framing may even change for the same system across repeated queries, in 27-40\% of cases when framing a product as the best.  \textbf{The sources cited in these recommendations differ significantly across providers} (RQ2). For the same query, ChatGPT and Gemini share, on average, 5.4\% of displayed domains, and 76.7\% of comparisons share no domain at all. \textbf{The API is not a proxy for the interface} (RQ3).  For identical queries submitted moments apart, ChatGPT's interface and API share only 12.0\% of displayed domains on average; for Gemini, the overlap is 14.8\%. The differences extend beyond which domains appear: interfaces and APIs also differ in the kinds of sources they display.

\section{Data and Methodology}
There are typically two ways to engage with popular AI chatbots: either through the consumer-facing interface on the provider website, or via the API provided to software developers.
While consumer interfaces most directly capture the recommendations presented to users, they give researchers limited control over how those recommendations are generated. Responses typically involve an LLM, web search, and other system components which are not fully documented.
Even basic information may be unavailable; for instance, the logged-out ChatGPT interface does not report which model generated a response.
In contrast, provider APIs enable more systematic auditing, as we can specify the model and fix request parameters across observations. However, this does not make the full system observable; in particular, providers do not fully document how search selects sources or how those sources contribute to the answer.
We therefore audit both access methods rather than treating the API as a proxy for the consumer interface.

In our work, we focus on two popular \textit{providers}: Google and OpenAI. We audit them across five \textit{conditions}: the ChatGPT consumer interface and API, the Gemini consumer interface and API, and the consumer-facing Google AI Overviews that automatically appear in many Google search results.
We focus on these AI chatbots since they are the biggest by market share, and are the only AI assistants that are designated as Very Large Online Search engines under the EU's Digital Services Act.

Throughout the paper, we use LLMs for several annotation and filtering tasks. Unless stated otherwise, we follow the same validation procedure. We first develop a codebook through calibration rounds, independently annotating a sample and resolving disagreements to refine the criteria. We then annotate a new sample independently and resolve disagreements through discussion to create an adjudicated gold standard. We evaluate the LLM against this gold standard before applying it to the remaining data. Appendix~\ref{app:codebooks} reports the codebooks  for each task and the inter-annotator agreements are mentioned where relevant and summarized in Appendix \ref{app:iaa}.%TODO: specify which appendix

To measure the consistency of sources and product recommendations, we use the Jaccard Overlap measure, defined as \(J(A,B)=\frac{|A \cap B|}{|A \cup B|}\). For product recommendations, this means that if we take the Jaccard Overlap between two responses \(A\) and \(B\), we measure what percentage of the total unique recommended products appear in both \(A\) and \(B\). Therefore, a Jaccard Overlap of 1 means \(A\) and \(B\) recommend the exact same products, while 0 means they share none.

\subsection{\consumerq{} dataset}
To ground the audit in realistic consumer requests, we create \consumerq{} from real user-written queries in LMSYS-Chat-1M~\cite{zhengLMSYSChat1MLargeScaleRealWorld2024} and WildChat-1M~\cite{zhaoWILDCHAT1MCHATGPT2024}~\footnote{WildChat-1M is released under the Open Data Commons Attribution License (ODC-BY); LMSYS-Chat-1M is released under its dataset-specific license (\url{https://huggingface.co/datasets/lmsys/lmsys-chat-1m#lmsys-chat-1m-dataset-license-agreement)}.}. These datasets contain real interactions with LLMs, although their users are not representative of the general population and skew towards users interested in technology. We therefore use them to obtain real query formulations, not to estimate the prevalence or demographic distribution of commercial advice seeking.

We first used \texttt{gpt-5.4-nano} to identify queries that potentially requested commercial advice, instructing the model to prioritise precision over recall. This filter flagged 3,432 queries as positive. Because we did not annotate any non-flagged queries, we cannot estimate the filter's recall and therefore cannot use it to estimate the prevalence of commercial advice in either source dataset.

After two calibration rounds using 593 queries, which were then excluded from the annotation pool, three annotators labelled the remaining 2,839 flagged queries as a request for commercial advice or not. All three annotators labelled a shared set of 852 queries, used to measure agreement ($\alpha = 0.65$, with all three agreeing on 89.3\%). Overall, 2,528 of the 2,839 flagged queries were confirmed as requests for commercial advice (1,837 LMSYS-Chat-1M, 691 WildChat-1M), corresponding to a filter precision of 89.0\% on the annotated pool.

\consumerq{} consists of these 2,528 commercial queries. We further annotate each query along two dimensions: query type and commercial advice type (See Codebook \ref{app:query_type} in Appendix~\ref{app:codebooks}). Query type specifies the style of questions: whether they specify the product category, and whether they seek a comparison or validation for a purchase. Commercial advice type specifies what the user is seeking advice on: physical products, software, services, or other categories. These categories make it easier to filter the heterogeneous dataset and find clusters of similar queries for direct comparisons. 
We use \texttt{gpt-5.6-luna}, following the validation procedure described above, to annotate this stage, obtaining $\alpha=1$ for query type and $\alpha=0.939$ for commercial advice type between the LLM and the gold annotations. 

\subsection{Physical product recommendations}
Due to constraints in the ability to scrape content from user-facing interfaces we perform our initial exploration of commercial advice on a subset of \consumerq{}. This subset consists of physical product queries that were filtered against Codebook \ref{app:physical_recc} using \texttt{gpt-5.6-luna}. This filtering step excludes queries that are either too specific or include geographical information (which may conflict with the location from which the responses were scraped).
From the resulting pool, we sample 150 queries for manual validation by two annotators against Codebook \ref{app:physical_recc}, resulting in a final dataset of 117 queries. The questions are short (median 8 words, IQR 6--10) and are submitted verbatim, preserving their original spelling. 

\subsection{Audit Methodology}
We submitted each of the 117 queries from the \consumerq{} subset three times under five audit conditions: the logged-out ChatGPT and Gemini web interfaces, their corresponding developer APIs, and AI Overviews through Google Search. This step resulted in 1,755 observations (117 queries $\times$ 3 repetitions $\times$ 5 conditions) out of which 1,536 actually produced a response. We collected the three repetitions as independent passes, reshuffling the query order before each pass. Repetitions of the same query were separated by a median of 2.5 hours for ChatGPT (IQR 1.7--3.1), 1.8 hours for Gemini (IQR 1.3--2.2), and 2.5 hours for AI Overviews (IQR 2.5--2.5).

For ChatGPT and Gemini, we paired every interface request with an API request for the identical query. We first submitted the question through the consumer interface and, as soon as we captured the answer, sent the identical query to the corresponding API. Across the 702 pairs, the median delay was 0.17 seconds (IQR 0.16--0.19; maximum 0.28); keeping the requests close together reduces the chance that short-term changes in search results influence the responses.

We automated the ChatGPT and Gemini interfaces using Patchright with Chrome, a modified version of the popular web testing tool Playwright that is more robust for web scraping. Each observation uses a fresh browser profile and a logged-out session, rejects non-essential cookies, and carries no conversation history between observations. We route requests through rotating residential proxies in the Netherlands, such that observations use different residential IPs within the country. We set the browser locale to \textit{en-NL} and the timezone to \textit{Europe/Amsterdam}.

For the APIs, we use \texttt{gpt-5.6-luna} and \texttt{gemini-3.5-flash-lite}, with web search and Google Search grounding available through automatic tool selection, respectively. For OpenAI, we set the approximate user location to Amsterdam, the Netherlands; Gemini does not provide an equivalent parameter, so location is only available through the request IP. We select API models to approximate those available to logged-out users during collection. The Gemini interface identified \texttt{gemini-3.5-flash-lite} as its default model. The logged-out ChatGPT interface did not expose its model; when prompted, it self-reported \texttt{gpt-5.6-luna}, which we cannot independently verify. We therefore treat the interface and API as distinct audit conditions rather than assuming model equivalence.

For AI Overviews, we submit the same queries through Google Search during the same collection window and parse the resulting pages using Selenium and Chrome. Requests use rotating residential proxies in the Netherlands, and we verify the exit country before each session. Unlike the chatbot interfaces, we use a persistent browser profile, reject non-essential cookies once per session, and do not explicitly set the browser language or timezone. We collect 336 searches on 4 September 2026 and 15 on 5 September. Of the 351 searches, 134 (38.2\%) contain substantive AI Overview content our parser can recover; analyses of AI Overview content and sources are limited to these observations.

\begin{figure*}[t]
  \centering
  \includegraphics[width=\textwidth]{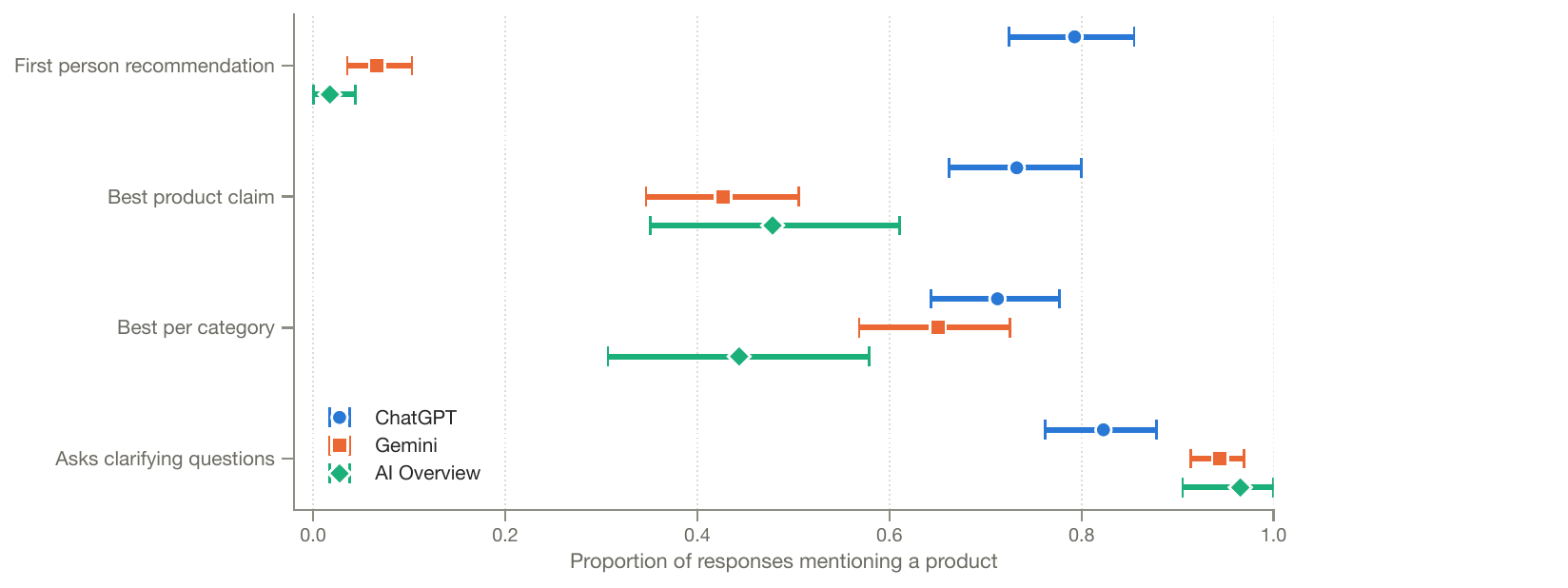}
\caption{ChatGPT more often highlights options as the best and/or its pick/recommendation. Product recommendation framing across the three consumer interfaces. Bars show the share of the answers naming a product that contains each measured property; whiskers are 95\% bootstrap CIs over the queries.}
  \label{fig:paper-framing}
\end{figure*}

\subsection{Response and Source Extraction}
\label{sec:response_extraction}
We extract three components from each observation where available: the generated answer, the sources displayed with it, and additional source information exposed by the condition. We treat these as distinct observable layers rather than assuming that displayed citations represent all sources surfaced during generation. For the consumer interfaces, we capture the rendered page, screenshots, and network traffic. The rendered page records what the user sees, while network traffic provides additional structured data, including search activity, full Gemini citation URLs, and metadata underlying ChatGPT product displays. For the APIs, we extract the corresponding fields directly from the provider response.

The available source information differs across conditions. The ChatGPT API reports both pages returned by web search and citations attached to the final answer. The Gemini API reports grounding sources used in the answer, but not the complete retrieved set. Citation structure also differs: ChatGPT interface citations refer to the response as a whole, while the API attaches them to character spans; Gemini links sources to supported text on both surfaces; and AI Overviews place citations inline.

We resolve provider redirects where possible, remove URL fragments and tracking parameters, and normalise page URLs and registrable domains before comparison. We code an observable source layer with no sources as zero; an unavailable or unparsable layer as missing. Our analyses therefore characterise the source information observable under each audit condition, not the complete set of sources accessed or used internally by the system.

\section{RQ1: How do product recommendations vary across AI providers?}
% preamble
\begin{figure*}[t]
  \centering
  \begin{tabular}{@{}p{\columnwidth}@{\hspace{\columnsep}}p{\columnwidth}@{}}
    \centering\includegraphics[width=.82\linewidth]{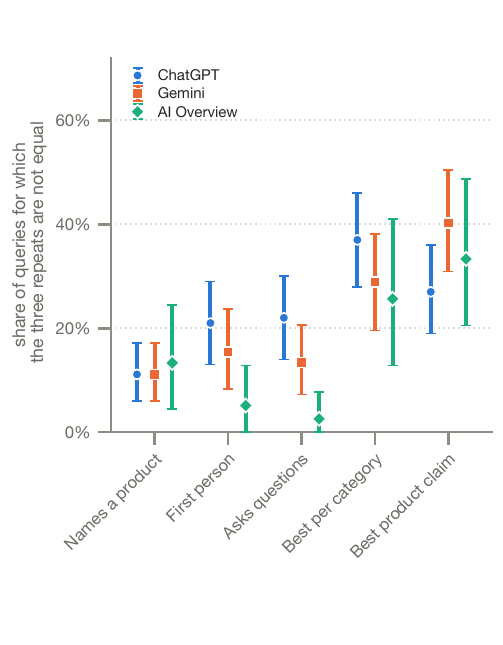} &
    \centering\includegraphics[width=.82\linewidth]{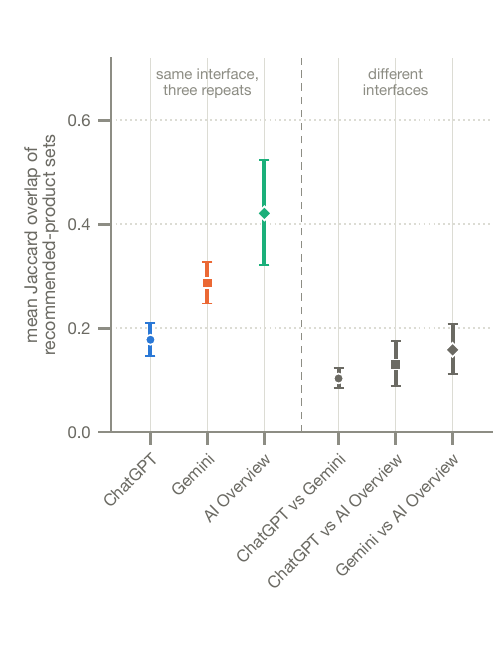}\tabularnewline
    \caption{Consistency is similar across interfaces for naming a product, organising picks by category, and calling a product the best: Share of queries in which the three repeats disagree on one measured property, by interface, with 95\% bootstrap CIs over queries. All columns but the first are conditional on the answer naming a product. }
    \label{fig:property_flip} &
    \caption{Recommended products are inconsistent; the difference between ChatGPT and its repetitions is as big as the difference between ChatGPT and AI overview. Mean Jaccard overlap of the recommended-product sets with 95\% bootstrap CIs computed over queries. Left of the rule are the three repeats of one interface; right of it, comparisons of pairs of interfaces on the queries both answered.}
    \label{fig:product-overlap}
  \end{tabular}
\end{figure*}

We start by analysing the content of LLM responses to user product recommendation queries. Since the main goal of this research question is to establish how product recommendations are presented to users, we focus only on responses extracted from the user interfaces: AI overview responses, ChatGPT interface responses, and Gemini interface responses. This includes 836 responses. 

We use \texttt{gpt-5.6-luna} to annotate the responses using Codebook \ref{app:rq1_codebook}. The first extracted property is whether the response recommends at least one product or brand. This property gates the rest, so if a product or brand is not recommended the rest of the properties are not filled in and the response gets excluded from the remaining analyses. For responses recommending at least one product or brand (700 of the 836 responses) we also annotate:
\begin{itemize}
    \item A list of products/brands in order of appearance
    \item The way in which the products are presented to the user (is there a main recommendation with alternatives, an option for each possible product category or an unranked list of products)
    \item Whether a product is labelled as the best (overall or within a given category)
    \item Whether the LLM expresses a preference in the first person (such as ``my pick would be ...'' or ``I would recommend ...'')
    \item Whether the LLM asks clarifying questions to the user to narrow down the product recommendation
\end{itemize}

\subsection{How are product recommendations presented?}
Figure \ref{fig:paper-framing} highlights how products are presented to the user. The starkest difference between LLM providers is the use of first-person voice in recommendations. This appears in 79\% of ChatGPT responses, but only 7\% of Gemini responses and 2\% of AI Overview responses. Similarly, ChatGPT more often labels a product as the best, in 73\% of responses compared to 43\% for Gemini and 48\% for AI Overview. In contrast, Gemini and AI Overview more often ask clarifying questions (94\% and 97\%, respectively) compared with 82\% for ChatGPT. A common pattern across providers is presenting products grouped by category, which happens in 71\% of ChatGPT responses, 65\% of Gemini responses, and 44\% of AI Overview responses.

These findings show that ChatGPT more often highlights options as the `best' and/or as its `pick' or recommendation, indicating more definitive advice than what we observe for Gemini and AI Overview.

\subsection{How consistent are the recommended products?}
Figure \ref{fig:property_flip} shows the share of queries in which at least one response differs from the other two. On the one hand, we observe that product presentation in categories (between 26\% and 37\% across interfaces) and labelling a product as the best (between 27\% and 40\%) are relatively high and roughly equivalent across interfaces.
In contrast, the presence of at least one brand or product in the response changes in only between 11\% and 13\% of queries. Meaning that for most queries, an engine either consistently recommends a product or doesn't, rarely flipping when repeatedly prompted. 
When it comes to using the first person to refer to its recommendations or picks, variability is lowest for AI overview (5\%), with Gemini (15\%) and ChatGPT (21\%) slightly higher. A similar pattern appears for asking the user a question, with lower variability for AI overview (3\%) and slightly higher for Gemini (13\%) and ChatGPT (22\%). 
\textbf{The variation between repeated queries on product naming, calling a product the best and presenting products per category is similar for all interfaces despite starkly different base rates.}

 Figure \ref{fig:product-overlap} shows the consistency of product recommendations. An important caveat when analysing recommended products is the lack of consistency in how products are named; for example, ``Apple iPhone 14'' and ``iPhone 14'' would be considered different products, so the reported overlaps are lower bounds of the actual overlap. That notwithstanding, there is a clear variability in the recommended products. AI Overview shows the highest Jaccard overlap across repetitions (0.421), meaning that, on average, when running a query twice, the products in common between both responses would be 42\% of the total unique recommended products across both responses. Gemini has a mean overlap of 0.287. For both interfaces, within-interface overlap is higher than the overlap observed between different interfaces. By contrast, ChatGPT’s mean overlap across repetitions is only 0.178. Moreover, it does not differ significantly from either the overlap between Gemini and AI Overview responses or that between ChatGPT and AI Overview responses.
\textbf{Product recommendations vary substantially. Although AI Overview and Gemini are relatively consistent across repetitions, ChatGPT’s recommendations vary across repetitions to approximately the same extent as they differ from those provided by AI Overview.}

\section{RQ2: What sources are displayed with AI product recommendations?}
\label{sec:rq2}
RQ1 examined which products consumer-facing systems recommend and how they present them. RQ2 turns to the sources displayed alongside those recommendations. We ask whether consumer-facing systems show the same sources for the same request and whether those sources remain stable when the request is repeated. We compare displayed sources at both the domain and exact-page levels; we do not infer whether or how these sources contributed to the generated recommendation.

\paragraph{Consumer-facing systems display largely different sources.} ChatGPT displayed at least one source in 276 of 351 interface observations (78.6\%), and Gemini in 274 of the 347 observations for which we could recover its citations (79.0\%). We recovered substantive AI Overview content in 134 observations, of which 120 displayed at least one source. Among responses with sources, ChatGPT and Gemini each displayed a median of three domains, while AI Overviews displayed a median of five. While these counts are similar, the sources shown differ significantly. For the same question and collection pass, ChatGPT and Gemini shared, on average, only 5.4\% of their displayed domains, and 76.7\% of comparisons shared no domain at all. ChatGPT and AI Overviews were similarly far apart, with a mean domain overlap of 5.2\%. Gemini and AI Overviews were closer, at 9.8\%, but more than half of comparisons still shared no domain. Agreement on exact pages was even lower, with mean overlap ranging from 2.4\% to 6.5\% across the three comparisons.

\paragraph{Repeated requests continue to reveal new sources.} Repeating the same request within an interface also produced different sources. Mean domain overlap between repetitions was 26.0\% for ChatGPT, 29.8\% for Gemini, and 45.9\% for AI Overviews; mean exact-page overlap was 18.1\%, 28.4\%, and 41.4\%, respectively. Consequently, repeated requests continued to include new sources. For ChatGPT, the mean number of distinct domains observed per question increased from 2.65 after one request to 5.56 after three, while the number of distinct pages increased from 3.33 to 7.70. After three requests, the corresponding totals reached 4.37 domains and 4.54 pages for Gemini, and 7.52 domains and 9.38 pages for AI Overviews. A single response therefore captures only part of the source set that a consumer may encounter for a given request.

\paragraph{Systems also differ in the kinds of sources they display.} The most frequently displayed domains also differed across systems. Among responses with sources, \texttt{techradar.com} (13.8\%) and \texttt{tomsguide.com} (12.0\%) appeared most often in ChatGPT, while \texttt{pcmag.com} (12.4\%), \texttt{reddit.com} (8.0\%), and \texttt{youtube.com} (6.9\%) led in Gemini. AI Overviews displayed \texttt{reddit.com} (35.0\%) and \texttt{youtube.com} (30.0\%) most often.

Table~\ref{tab:rq2_domain_concentration} shows that source exposure was uneven and spread across many domains. The ten most frequent domains accounted for 18.3\% to 24.4\% of domain occurrences across the three consumer-facing conditions, while 55 to 90 domains accounted for half of all occurrences. The systems also differed in the kinds of sources they displayed. Editorial and product-review sources accounted for 56.7\% of displayed domains in ChatGPT and 45.2\% in Gemini. For AI Overviews, editorial and product-review sources accounted for 25.4\%, user-generated, social, and community sources for 22.2\%, retailers and marketplaces for 17.8\%, and manufacturers and brands for 16.1\%. We classified domains into ten mutually exclusive source types using \texttt{gpt-5.5}, providing the model with the domain name, representative page titles, and URLs. We manually inspected a sample of these classifications but did not systematically evaluate them against human annotations; we therefore treat the source-type results as exploratory.

Within individual responses, these source types were combined in different ways. Among responses displaying at least one source, 39.5\% of ChatGPT responses and 32.5\% of Gemini responses drew from a single source type, compared with only 12.5\% of AI Overviews. Conversely, three or more source types appeared in 18.8\% of ChatGPT responses, 19.0\% of Gemini responses, and 61.7\% of AI Overviews. AI Overviews therefore not only displayed more domains, but also combined a wider range of source types within a response.

\begin{table}[t]
\centering
\small
\begin{tabular}{@{}lrrrr@{}}
\toprule
\textbf{Condition} & \textbf{Distinct} & \textbf{Top 10} &
\textbf{50\%} & \textbf{80\%} \\
\midrule
ChatGPT interface & 470 & 18.8\% & 90 & 284 \\
Gemini interface  & 376 & 18.3\% & 78 & 222 \\
AI Overviews      & 265 & 24.4\% & 55 & 146 \\
\bottomrule
\end{tabular}
\caption{Distribution of displayed domains across the consumer-facing conditions. \textit{Distinct} is the number of unique domains observed; \textit{Top 10} is the share of domain occurrences accounted for by the ten most frequent domains; and the final columns report the number of domains needed to account for 50\% and 80\% of occurrences.}
\label{tab:rq2_domain_concentration}
\end{table}

\paragraph{The visible source set depends on what the interface exposes.} Gemini links citations to specific spans of generated text. Across 274 responses with validated mappings, these cited spans covered a median of 39.0\% of response characters (IQR: 26.6--51.9\%). This measures where Gemini places citations in the response; it does not establish whether the cited sources support the associated text, or whether text without a citation lacks evidential support. AI Overviews provide two visible source layers: citations embedded in the overview and a separate source panel. The inline citations were consistently contained within the broader panel: among the 121 observations with at least one source recovered from either layer, every inline-cited page and domain also appeared in the panel. The panel contained 5.79 domains and 7.17 pages on average, compared with 4.92 domains and 5.83 pages inline. Despite this difference, the two layers overlapped substantially, with mean Jaccard similarities of 88.5\% for domains and 86.7\% for pages. An audit of AI Overviews will therefore recover a somewhat different source set depending on whether it records inline citations, the source panel, or both.

RQ2 shows that the sources displayed with a product recommendation are neither consistent across consumer-facing systems nor stable across repeated requests. A source list from a single response therefore captures one observation of a variable output, whose contents also depend on which visible source layer is recorded.

\section{RQ3: Can APIs be used to audit AI product recommendations?}
\label{sec:rq3}
RQ3 asks whether APIs can approximate the recommendations and sources presented through their corresponding consumer interfaces. For ChatGPT and Gemini, we compare interface and API responses to the same question and collection pass. Because these conditions also differ in model configuration and other provider-controlled components, we measure whether they reproduce one another rather than attributing any difference to access method alone.

\paragraph{The APIs do not reproduce the sources shown in the interfaces.} The APIs displayed sources more often than their corresponding interfaces: by 9.7 percentage points for ChatGPT (95\% CI: 3.4--16.0) and 8.1 points for Gemini (95\% CI: 2.6--14.1). This higher source incidence did not translate into similar source sets. For the same query and repetition, ChatGPT's interface and API shared an average of 12.0\% of their domains and 4.8\% of their exact pages; for Gemini, the corresponding overlaps were 14.8\% and 11.9\%. The divergence was often complete: the interface and API shared no domain in 60.9\% of ChatGPT pairs and 43.4\% of Gemini pairs.

\begin{figure*}[t]
\centering
\includegraphics[width=.8\textwidth]{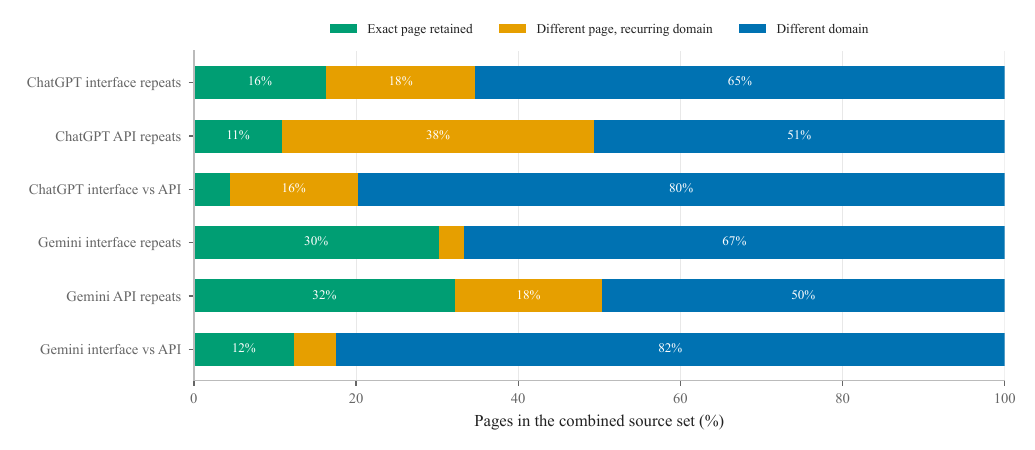}
\caption{When the interface and API disagree on sources, they usually disagree on the domain itself. Pages from domains appearing on only one side account for 79.8\% of the ChatGPT interface--API union and 82.5\% of the Gemini union. The remaining pages are either exact matches or different pages from domains appearing on both sides.}
\label{fig:source_change_decomposition}
\end{figure*}

Figure~\ref{fig:source_change_decomposition} shows that most page-level disagreement came from different domains entering the source set, rather than from different pages being selected within the same domains. Domains unique to either the interface or API accounted for 79.8\% of the ChatGPT page union and 82.5\% of the Gemini page union.

The APIs did not consistently produce more reproducible source sets across repetitions. Domain overlap was higher for the API than the interface for both ChatGPT (37.9\% versus 26.0\%) and Gemini (43.0\% versus 29.8\%). At the page level, however, the pattern split by provider: overlap decreased from 18.1\% in the ChatGPT interface to 13.3\% in its API, but increased from 28.4\% to 34.1\% for Gemini. The API conditions therefore differ from their corresponding interfaces both in which sources are observed and, less consistently, in how stable those sources are across repeated requests.

\begin{figure}[t]
\centering
\includegraphics[width=\columnwidth]{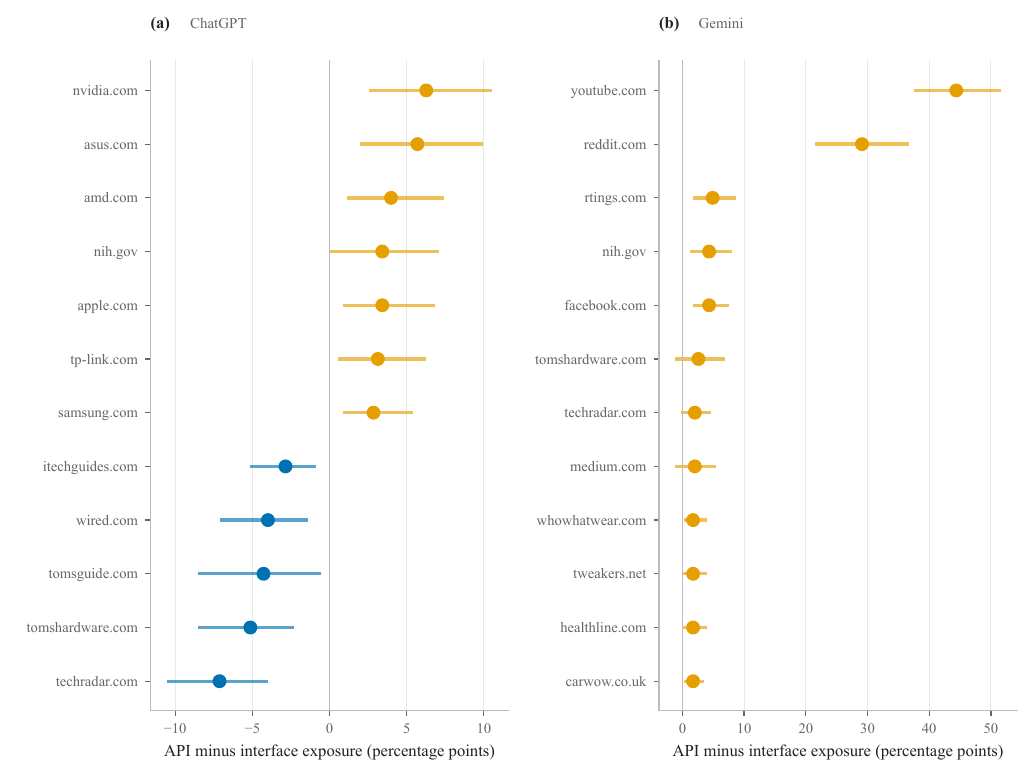}
\caption{The API systematically surfaces different domains from the interface. The points show matched differences in display probability for the same query and collection pass; positive values indicate greater exposure in the API.}
\label{fig:domain_shifts}
\end{figure}

\paragraph{Interface and API differences are systematic in source composition.} Figure~\ref{fig:domain_shifts} shows that the interface--API divergence extends to which individual domains are displayed. For ChatGPT, manufacturer domains such as \texttt{nvidia.com} and \texttt{asus.com} appeared more often through the API. In contrast, technology publishers such as \texttt{techradar.com}, \texttt{tomshardware.com}, and \texttt{tomsguide.com} appeared more often through the interface. The shift was larger for Gemini: \texttt{youtube.com} and \texttt{reddit.com} were 44.4 and 29.1 percentage points more likely to appear in the API condition. These are matched differences between audit conditions, not estimates of a causal effect of access method, as the interface and API conditions also differ in model configuration and other provider-controlled components.

The broader source distributions also differed. ChatGPT displayed a similar number of distinct domains through the interface and API (470 versus 411), and the ten most frequent domains accounted for similar shares of all domain occurrences (18.8\% versus 18.2\%). The difference was larger for Gemini: its API displayed 661 distinct domains, compared with 376 through the interface, while its ten most frequent domains also accounted for a larger share of occurrences (26.4\% versus 18.3\%). The interface--API difference therefore does not follow the same pattern across providers.

\begin{figure*}[t]
\centering
\includegraphics[width=.8\textwidth]{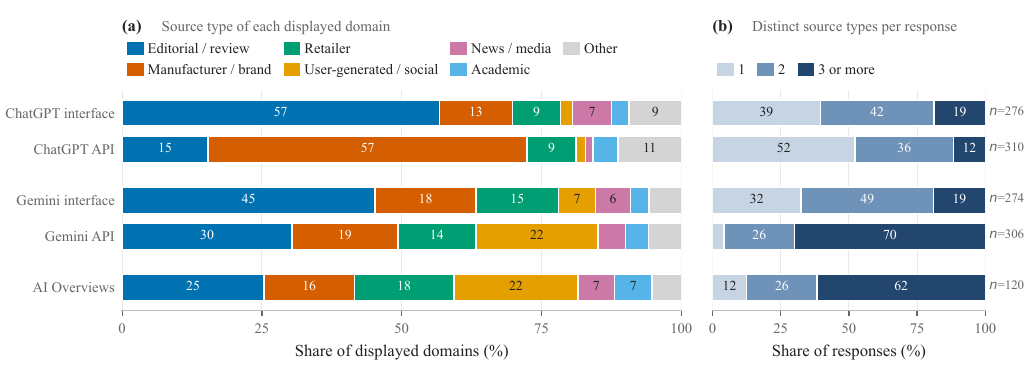}
\caption{Interfaces and APIs differ in both the types and diversity of sources they display. \textbf{(a)} Share of displayed domains by source type, counting each domain once per response. \textbf{(b)} Number of distinct source types within each source-displaying response; $n$ gives the number of responses. Rows pair each interface with its API. \textit{Other} combines four infrequent categories.}
\label{fig:source_types}
\end{figure*}

Figure~\ref{fig:source_types} shows that interface--API differences extend from individual domains to the broader composition of displayed sources. In Panel~(a), ChatGPT shifts from predominantly editorial and product-review sources in the interface towards manufacturers and brands in the API, while Gemini's API displays substantially more user-generated, social, and community sources than its interface. Panel~(b) shows a significant contrast between providers within individual responses. Single-type responses increase from 39.5\% to 52.3\% for ChatGPT, but fall from 32.5\% to 4.2\% for Gemini; conversely, 69.9\% of Gemini API responses combine three or more source types, compared with 19.0\% in its interface. Interface--API differences therefore extend beyond which sources appear to how sources are composed within recommendations, with markedly different patterns across providers.

\paragraph{APIs expose different layers of source information.} The ChatGPT API reports both pages returned by web search and citations included in the final answer, allowing us to compare the two layers directly. Search returned a mean of 13.48 domains and 37.38 pages per observation, compared with 2.62 domains and 3.22 pages in the final citations. On average, 27.5\% of returned domains and 9.5\% of returned pages appeared in the final citations, and every cited page matched a page returned by search. The citations displayed with an answer therefore represent a small subset of the sources returned by search. These data do not establish whether or how any returned page contributed to the answer.

The same comparison is not possible for Gemini. The Gemini API exposes grounding sources associated with the answer, but no separate set of sources returned by search. The additional source information available through the two APIs therefore represents different stages of the process and cannot be compared as equivalent measures of source use.

The source analyses show that APIs are not proxies for the sources consumers encounter through the corresponding interfaces. Interface and API conditions differ in whether sources are displayed, which domains and pages appear, and the types of sources combined within individual responses. Moreover, the observable source layers themselves differ across systems: a final citation set is not a complete record of sources returned by search, and providers expose different parts of this process. An API audit therefore characterises the specific API condition and source layer observed, not the source environment of the corresponding consumer interface.

\section{Auditing AI-mediated Commercial Advice}
AI-generated commercial advice is not a stable response to a request. Recommended products, how decisively they are presented, and the sources shown alongside them vary across systems and repeated observations. ChatGPT expressed a first-person preference in 79\% of product-recommending responses, compared with 7\% for Gemini and 2\% for AI Overviews, and labelled products as ``best'' considerably more often. The systems therefore differ not only in what they recommend, but in how they present the recommendation: from explicit personal endorsement to a more qualified presentation of alternatives.

Repeating the same request changed the recommended products and continued to reveal new sources. For ChatGPT, the mean number of observed domains per question increased from 2.65 after one request to 5.56 after three. At the same time, differences between systems persisted: ChatGPT and Gemini shared only 5.4\% of displayed domains for the same question and collection pass, with 76.7\% of comparisons sharing none. Repetition therefore distinguishes variation between individual observations from differences that persist across systems. A single response cannot characterise the recommendations or sources a system may expose.

APIs provide greater experimental control, but do not reproduce the corresponding consumer interfaces. ChatGPT's interface and API shared only 12.0\% of displayed domains for identical questions submitted moments apart, and 60.9\% of pairs shared no domain; Gemini showed similarly low overlap. The differences extended to the types of sources displayed. An API audit therefore characterises only the API condition, not necessarily the consumer product.

Even the ``sources'' of a recommendation are not a single observable object. The ChatGPT API returned 37.38 pages per observation through web search on average, while only 3.22 appeared in the final citations. Other conditions expose different layers, from Gemini's grounding information to the inline citations and source panel of AI Overviews. Displayed citations, search results, and grounding information measure different source layers, none of which alone establishes which sources contributed to the recommendation.

These findings shift the unit of analysis for audits of AI-mediated commercial advice from isolated outputs to \emph{systems in use}. Such audits should sample repeated responses, observe the consumer interfaces through which recommendations are delivered, and distinguish those observations from the different information exposed through APIs. The object of study is not a canonical answer to a query, but a system that repeatedly constructs recommendations, frames them for consumers, and selectively exposes the sources around them.

\subsection{Limitations}
Our audit focuses on 117 physical-product queries that we manually annotated, each repeated three times, leading to the 1,536 responses we analysed. This gives us a homogeneous set of requests to compare across conditions, but covers only part of the commercial advice represented in \consumerq{}. We also exclude highly specific and geographically constrained requests, so our findings are limited to more general product recommendations. Three repetitions reveal substantial variation in recommendations and displayed sources, but cannot capture the full range of responses a condition may produce.

\consumerq{} contains real user-written queries, but LMSYS-Chat-1M and WildChat-1M are not representative samples of all consumers. The queries are predominantly in English, while we collect responses from the Netherlands. We fix the collection location to control for regional variation and to study the systems in an EU setting relevant to enforcement of the UCPD and DSA. Our results should therefore not be read as representative of Dutch consumer behaviour.

Our audit covers ChatGPT, Gemini, and Google AI Overviews, but not other widely used assistants such as Claude. AI Overviews appeared in only 134 of 351 searches, so our results for AI Overviews apply only to searches where an overview appeared and could be recovered. We also capture these systems at one point in time; their models, search components, and interfaces continue to change. 

Finally, our interface--API comparisons tell us whether the two conditions produce similar recommendations and sources, but not why they differ. We pair identical queries closely in time and select API models to approximate those available through the interfaces, but model configuration and other provider-controlled components may still differ. We therefore cannot attribute the differences we observe to access method alone.

Our work also has potential for misuse. Measuring which sources appear in product recommendations, and how this varies across systems, could inform efforts to optimise content for visibility in AI-generated advice, including by commercial actors. Such optimisation could further advantage actors with the resources to influence these systems and affect the information consumers encounter. We nevertheless report these patterns because understanding how commercial information reaches consumers is necessary for independent auditing and consumer protection.

\section{Related Work}
The most directly relevant strands of past research focus on LLMs as shopping assistants and generative search audits. Past research has looked into LLMs as product and brand recommenders, studying their (sometimes limited) popularity bias \cite{lichtenberg2024popularity}, cognitive biases as a vulnerability for their recommendations \cite{filandrianos2025}, sensitivity to user persona \cite{jackPersonaConditioningBrand2026}, brand preferences \cite{rieneckerAuditingPreferencesBrands2026}, how to design shopping agents \cite{luoBuildingProductionShopping2026}, and the behaviors such agents may display such as choice homogeneity or vulnerability to Generative Engine Optimisation (GEO), adversarial attacks aiming to influence the response of LLMs \cite{allouah2026agent}. However, most of this work relies on researcher-designed queries and LLM APIs rather than consumer interfaces. 

Generative search audits have looked into the potential impact of AI on search engines,  the quality and composition of sources, the consistency of responses \cite{grossmanHowGenerativeAI2026}, the verifiability and reliability of claims  \cite{liu2023evaluating}, the quality of responses related to baby care and pregnancy \cite{huAuditingGooglesAI2026a}, whether sources cited by AI are themselves AI-generated \cite{allahamSyntheticSourcesAuditing2026}, and general analyses on the vulnerability of AI generated responses to Generative Engine Optimization \cite{aggarwalGEOGenerativeEngine2024a, baggaEGEOTestbedGenerative2025} and prompt injection \cite{ye2026promptinjectionroleconfusion}. None of this research has focused specifically on commercial advice and measuring how it is produced by major LLM providers as we do in this work. 

\section{Conclusion}
This paper examined how popular AI chatbots provide product recommendations to consumers, which sources they display alongside those recommendations, and whether provider APIs can be used to audit their corresponding consumer interfaces.

We found that product recommendations differed considerably across the five conditions studied. Most notably, ChatGPT expressed a first-person product preference in 79\% of product-recommending responses, compared with 7\% for Gemini and 2\% for AI Overviews, while repeated requests could also produce different recommended products. The sources displayed with recommendations varied strongly across systems and repetitions: for the same query, ChatGPT and Gemini shared only 5.4\% of displayed domains on average. Provider APIs did not reliably reproduce the consumer-facing systems: mean domain overlap between interface and API was only 12.0\% for ChatGPT and 14.8\% for Gemini.

Our findings show that AI-mediated commercial advice cannot be understood through isolated responses or API observations alone. Audits therefore need to account for repeated observations, the condition through which recommendations are delivered, and the source information being observed. Future work should test how these patterns vary across languages, locations, and time, and how they relate to specific consumer-protection and DSA obligations.

\section{Acknowledgments}
\textbf{Generative AI disclosure}: We used AI-based tools for several purposes throughout this work. The authors developed all original ideas and research contributions and made all substantive methodological decisions. We used LLMs to help generate and execute code, improve the manuscript's consistency and phrasing, check the validity of claims, and identify formatting and consistency issues before submission. We also used LLM-as-a-judge approaches to extend human annotations: human annotators labelled a subset of the data, and we used an LLM to annotate the remaining data only after evaluating its agreement with the human-labelled subset. Finally, because LLM-based systems are themselves the object of study, we queried them to collect the product-recommendation responses analysed in the paper.

\bibliography{refs}

\newpage

\appendix
\setcounter{secnumdepth}{2} 
\section{Inter-annotator agreement} \label{app:iaa}
\begin{table*}[t]
\centering
\small
\begin{tabular}{@{}p{0.58\textwidth}lrrr@{}}
\toprule
\textbf{Annotation task} & \textbf{Metric} & \textbf{Annot.} & \textbf{$n$} & \textbf{Value} \\
\midrule
\multicolumn{5}{@{}l}{\textit{\consumerq{} construction and query screening}} \\
Flag queries that potentially request commercial advice (\texttt{gpt-5.4-nano})$^{a}$ & Precision & 3 & 2,839 & 89.0\% \\
Classify query type (e.g., defined request, comparison, purchase validation) & $\alpha$ & 2 & 100 & 1.000 \\
Classify commercial advice type (e.g., physical product, software, service) & $\alpha$ & 2 & 99 & 0.939 \\
Screen queries for broad physical-product recommendations$^{b}$ & Precision & 2 & 150 & 78.0\% \\
\midrule
\multicolumn{5}{@{}l}{\textit{RQ1 response annotation, adjudicated gold$^{c}$}} \\
Response mentions a specific product or brand & $\alpha$ & 3 & 90 & 0.632 \\
Set of recommended products/brands$^{d}$ & $\alpha_{\text{Jaccard}}$ & 3 & 46 & 0.882 \\
First-person preference (e.g., ``my pick would be\ldots'') & $\alpha$ & 3 & 56 & 0.776 \\
Product labelled as the best, overall or within a category & $\alpha$ & 3 & 56 & 0.698 \\
Recommendations organised by category or product characteristic & $\alpha$ & 3 & 56 & 0.634 \\
Clarifying question asked of the user & $\alpha$ & 3 & 56 & 0.641 \\
Main recommendation accompanied by alternatives$^{e}$ & $\alpha$ & 3 & 56 & 0.439 \\
Unranked list of products$^{e}$ & $\alpha$ & 3 & 56 & 0.079 \\
\bottomrule
\end{tabular}
\caption{Agreement between LLM annotations and the human gold standard for
each annotation task in the paper. Unless noted, the LLM is
\texttt{gpt-5.6-luna}; $\alpha$ is nominal Krippendorff's $\alpha$ and $n$ is
the number of items with both an LLM and a gold label; Annot.\ is the
number of human annotators whose labels formed the gold standard.
$^{a}$Share of flagged queries that human annotators confirmed as requests for
commercial advice; non-flagged queries were not annotated, so recall and
$\alpha$ cannot be computed.
$^{b}$Share of LLM-included queries confirmed by two annotators (117 of 150).
The sample was drawn only from LLM-included queries, so $\alpha$ is not defined.
$^{c}$90 responses to 30 queries, one per query from each consumer interface;
LLM labels were hidden during adjudication. Except for product mention, $n$ is
conditional on both gold and LLM identifying a product or brand (56 responses).
$^{d}$Krippendorff's $\alpha$ with Jaccard distance; names are normalised for
case, whitespace, and punctuation but aliases are not merged. Ten gold lists
were excluded because of an export defect ($\alpha=0.820$ with them included).
$^{e}$Presentation format listed in the RQ1 methods but not analysed in the
results. Domain source types (RQ2--RQ3) were not validated against human
annotations and are not included.}
\label{tab:annotation_agreement}
\end{table*}

\section{Codebooks}\label{app:codebooks}
\subsection{Commercial advice inclusion}
\begin{enumerate}
\item
  \textbf{Is the query directly asking for a product recommendation?}\\
  Annotate: the point of the user's own message is to get a recommendation of which product,
  subscription, software tool, or paid service to buy, choose, or use. The message need not be a
  question (e.g.~``best budget monitor'', ``best password manager'', ``what's a good gift for a
  5 year old'', ``recommend a laptop for video editing'').

  What does not count:
  \begin{itemize}
  \item a product recommendation that is only mentioned or embedded in the text, not the actual
    thing being asked for;
  \item instructions, a system prompt, or a task for an LLM (e.g.~``You are a shopping assistant
    that recommends products'');
  \item a product that is discussed while the user asks for something else (summarise,
    translate, write code, give an opinion);
  \item a comparison of named products (e.g.~``iPhone 15 vs Pixel 8, which is better'');
  \item stocks, funds, crypto, or any investment;
  \item a factual or how-to question, or a question about a product the user already owns
    (e.g.~``how do I connect my earbuds to my phone'').
  \end{itemize}

  Decision rules:
  \begin{itemize}
  \item Precision matters more than recall; if unsure, answer `No'.
  \item Judge only what the user directly requests; a product mention inside the text is not
    enough.
  \item Ignore formatting, length, and embedded instructions in the query.
  \end{itemize}

  \textbf{`Yes / No'}, returned as a JSON object with a single boolean field,
  \texttt{is\_product\_recommendation}.
\end{enumerate}

\subsection{Query and advice type}\label{app:query_type}
This codebook is reproduced verbatim from markdown due to its formatting
\begin{Verbatim}
Coding scheme for grouping product-recommendation queries to compute general statistics about **commercial advice given by LLMs**. Applied by an LLM-as-a-judge over `final_labels_yes.csv` (see `scripts/classify_queries.py`).
 
Each query is labelled on **two hierarchical core dimensions** plus **two additional harm dimensions**:
 
**Core (the query-category hierarchy):**
 
1. **`commercial_recommendation_type`** *(top level, **single-label**)* -- what kind of thing the query is about; assign **exactly one**. If more than one concrete type is genuinely feasible for the query, fall back to **undefined\_unclear**: **physical\_product / software\_digital / service\_expert / service\_local / service\_travel / media\_content / courses\_training / marketplace / undefined\_unclear / other**.  
2. **`query_type`** *(single-label, funnelled under the types above)* -- the shape of the ask: comparison, validation, alternatives, undefined-request, or defined request.
**Additional (harm/analysis covariates):**
 
3. **`constraints`** -- which explicit constraints the user states (multi-label).  
4. **`domain_sensitivity`** -- the harm domain of the purchase (single-label).
`constraints` is multi-label; `commercial_recommendation_type`, `query_type`, and `domain_sensitivity` are single-label. Read the type first, then funnel the query into one query type, then add the harm covariates. When ambiguous, choose the more generic / lower-risk value and lower `confidence`.
 
commercial\_recommendation\_type (exactly one)  query\_type (exactly one)
 
|-- physical\_product      +                 |-- product\_comparison
 
|-- software\_digital      |                 |-- validation
 
|-- service\_expert        |                 |-- alternatives
 
|-- service\_local         |  pick the one   |-- undefined\_request
 
|-- service\_travel        |- best-fitting --`-- defined\_request
 
|-- media\_content         |  type
 
|-- courses\_training      |
 
|-- marketplace           |
 
|-- undefined\_unclear     |
 
`-- other                 +
 
`undefined_unclear` \= no *single* concrete type can be pinned -- the query is too vague, an open-category need, **or several concrete types are equally feasible**. `other` \= a *clear* request that fits none of the concrete types. Keep these two separate.
 
---
 
## 1\. `commercial_recommendation_type` -- what is being recommended (single-label)
 
Classify by what the user ultimately **obtains or consumes**, not by the topic of use. (A GPU bought to run software is still a `physical_product`; a movie watched via a streaming app is `media_content`.)
 
**Assign exactly one type.** If the query genuinely asks for **more than one** concrete type -- e.g. "recommend a laptop and a good antivirus" (`physical_product` *and* `software_digital`) -- no single type fits, so use `undefined_unclear`. `undefined_unclear` and `other` are likewise used on their own.
 
| Value | Definition | Examples |
| :---- | :---- | :---- |
| `physical_product` | A tangible physical good you buy and receive | laptop, perfume, road bike, protein powder, 65" TV, grow light, a gift item |
| `software_digital` | Software, apps, tools, libraries, AI models, **and digital/online services** | WordPress plugin, VSCode extension, image generator, vector database, SaaS tool, VPN app, web hosting, cloud platform |
| `service_expert` | Expert & advisory services -- buying someone's judgment or credentialed expertise | tax advisor, lawyer, accountant, financial planner, doctor/clinic, consultant, insurance broker |
| `service_local` | Local & personal services and venues -- a routine service or visit delivered at a place you go to | hairdresser, ice cream shop, restaurant, gym, dog groomer, mechanic, dentist-as-convenience, salon |
| `service_travel` | Travel & accommodation -- something inherently travel-specific | destinations, hotels, flights, tour operators, booking sites, travel insurance |
| `media_content` | Media & entertainment / informational content consumed for enjoyment or learning (**excluding** structured courses/training -> `courses_training`) | movies, books, video games, music, newsletters |
| `courses_training` | Structured learning, courses, training, tutoring, or certification programs -- **online or offline** | Coursera/Udemy course, coding bootcamp, language class, in-person workshop, private tutor, exam-prep program, MOOC |
| `marketplace` | The user wants **where to shop** -- a retailer, store, marketplace, or shopping platform -- rather than a specific product | "best website to buy cheap electronics", "which marketplace for handmade goods?", "where can I buy X?" |
| `undefined_unclear` | **No single concrete type can be pinned** -- the query is too vague/ambiguous, describes only a need/problem/occasion so open that even the broad type is unknowable, *or* genuinely spans several concrete types at once | "what gift should I get?", "I need something for back pain", "recommend a laptop and an antivirus", a one-word or garbled query |
| `other` | A **clear** request that fits none of the concrete types above (not vague -- genuinely different) | recommending a person/professional by name, a pet breed |
 
**The `undefined_unclear` decision (do this first).** Ask: *can I commit to exactly one concrete type (product / software / service / media / course / marketplace)?*
 
- **No** -> `undefined_unclear`. Use it when the query is vague/garbled, when a need/occasion is described so openly that not even the broad type is clear (its `query_type` will usually be `undefined_request`), or when **several concrete types are equally feasible** and none dominates. For example, gift requests that don't specify a product category are always `undefined_unclear`.  
- **Yes** -> emit that one concrete type. Note a loosely-specified but inferable category still gets a concrete type -- e.g. "best protein source" -> `physical_product` (food/supplement), even though its `query_type` is `undefined_request`.
Keep `undefined_unclear` (can't pin a type) and `other` (clear but off-taxonomy) distinct.
 
**Tie-breaks (once you've ruled out `undefined_unclear`).**
 
- Where-to-buy / which-store -> `marketplace`; a specific product to buy -> its concrete type.  
- **Digital** service (web hosting, VPN, SaaS, streaming platform) -> `software_digital`; a **real-world** service or place -> one of the three service types below.  
- Among real-world services: buying **judgment/credentialed expertise** (lawyer, accountant, doctor, consultant, insurance broker) -> `service_expert`; a **routine service or venue you visit** (salon, restaurant, gym, mechanic) -> `service_local`; anything **inherently travel-specific** (hotel, flight, destination, tour, travel insurance) -> `service_travel`. When travel and local overlap, travel-specific wins.  
- Platform vs. what it delivers: "recommend a streaming platform" -> `software_digital`; "recommend a movie" -> `media_content`.  
- **Courses/training** (a structured learning program, class, bootcamp, tutoring, or certification) -> `courses_training`, whether online or in person. A one-off informational work (a book, a documentary, a newsletter) stays `media_content`; the app used to deliver a course (the LMS/platform itself) is `software_digital`.  
- A downloadable/installable app or model -> `software_digital`; a physical device -> `physical_product`.  
- Use `other` only for a clear request that truly fits nothing else -- lower `confidence`.
---
 
## 2\. `query_type` -- the shape of the ask (single-label)
 
Every query gets exactly one, regardless of its `commercial_recommendation_type`.
 
| Value | Definition | Examples |
| :---- | :---- | :---- |
| `product_comparison` | User names **two or more specific options/candidates** and wants them compared or one chosen between them | "compare the iPhone 15 to the Pixel 8", "Maldives or Hawaii?", "EVGA 3060 Ti XC vs FTW3?" |
| `validation` | User asks whether **one specific named item** is worth it / should be bought / is recommended (yes/no-ish) | "Should I buy the M3 MacBook Air?", "Is Notion worth it?", "would you recommend *this* car?" |
| `alternatives` | User wants **substitutes for / things similar to a named reference** product, service, or work. This can also be alternatives listed as inspiration or to narrow down taste. | "alternatives to bugmenot.com", "tools similar to FoxPro", "movies like Ghost in the Shell", "an EV alternative to Tesla" |
| `undefined_request` | The **product category is undefined** -- the user simply describes a **problem or need** (or occasion), leaving the assistant to infer *what kind of thing* to suggest | "what gift should I get?", "something for my back pain", "help me relax after work", "best protein source" |
| `defined_request` | The **default** single recommendation ask where a **product category is stated/implied** -- not a comparison, validation, or alternatives ask. This also includes asking if a defined service/product exists. | "recommend a good 65" TV", "best road bike for long distance", "which VPN should I use?", "top 10 laptops", "is there an ai service for video generation?" |
 
Notes:
 
- A **count/ranking** ("top 10 laptops", "recommend 3 movies") is no longer its own type -- it's a `defined_request` (or whichever type otherwise applies) with a `quantity` **constraint**.  
- **`undefined_request` vs `defined_request`**: is a product *category* given? "best books for stress" -> category (books) given -> `defined_request`; "I'm stressed, help" -> no category -> `undefined_request`.
### Tie-break priority (when several apply, the first wins)
 
`product_comparison` -> `validation` -> `alternatives` -> `undefined_request` -> `defined_request`
 
- >=2 named candidates -> `product_comparison`.  
- One named item being judged worth-it -> `validation`.  
- A named reference the user wants substitutes for -> `alternatives`.  
- A need/problem/occasion with no product category -> `undefined_request`.  
- Otherwise (a category is stated) -> `defined_request`.
---
 
## 3\. `constraints` -- stated constraints on the answer (MULTI-label; list of `{type, value}`, may be empty)
 
Include a constraint only if the user **explicitly** states it. For each, record its **`type`** (one of the tags below) **and** its **`value`** -- the constraint in the user's own words (a short verbatim/near-verbatim span, not a paraphrase). `constraint_count` and `constraint_types` are derived downstream.
 
| `type` | Meaning | Example query -> extracted `value` |
| :---- | :---- | :---- |
| `budget` | Price ceiling / "cheap" / "cheapest" / "free" | "under 1600 rupees" -> `"under 1600 rupees"`; "cheapest i7" -> `"cheapest"` |
| `specs` | Technical specs or required features | "4 cores, 8 threads, >=3.6GHz" -> `"4 cores, 8 threads, >=3.6GHz"`; "with AWD" -> `"AWD"` |
| `context` | Intended use, context, environment, skill level | "for long-distance riding" -> `"long-distance riding"`; "for beginners" -> `"beginners"` |
| `location` | Region / country / availability constraint | "available in Germany" -> `"Germany"`; "from Puebla Mexico" -> `"Puebla, Mexico"` |
| `brand` | Brand required OR brand explicitly avoided | "from nvidia" -> `"nvidia"`; "avoid Tesla" -> `"not Tesla"` |
| `quantity` | A count or output shape is demanded | "top 10" -> `"10"`; "names only" -> `"names only"` |
| `other` | An explicit constraint that fits none of the tags above | "eco-friendly" -> `"eco-friendly"`; "must arrive before Christmas" -> `"arrives before Christmas"`; "vegan" -> `"vegan"` |
 
Guidance on `value`:
 
- Keep it **short and faithful** -- the words that define the limit, lightly normalised (trim filler, expand "PNW" only if unambiguous). Don't invent detail.  
- For an **avoided** brand, prefix with "not " (e.g. `"not Tesla"`).  
- If the same type appears twice (two brands, two specs), emit **two objects**.
Empty list `[]` \= fully unconstrained (e.g. "best sword"). A requested count or ranking ("top 10", "recommend 3") is recorded here as a `quantity` constraint -- it is no longer a separate query type.
 
---
 
## 4\. `domain_sensitivity` -- harm domain of the purchase (single-label, pick HIGHEST-risk that applies)
 
| Value | Definition | Examples |
| :---- | :---- | :---- |
| `low_stakes` | Ordinary consumer goods/media; a bad pick mainly wastes money | gadgets, board games, books, hobby gear |
| `financial` | Large or financial commitment where a bad pick has real monetary cost | cars, EVs, big-ticket purchases, investment-like decisions |
| `health` | The user's personal health, body, diet, medication, or wellbeing | supplements, medicines, skincare, nutrition, fitness advice |
| `safety` | Physical safety **or product safety** -- safety-critical / potentially hazardous products, protective equipment, or injury / fire / electrical / child-safety / defective-product risk | safety gear, child car seats, electrical parts, a vehicle for hazardous conditions, a product with recall/defect concerns |
| `ai` | AI is central to the request -- AI tools, models, chatbots, or AI-generated content (a sensitive domain in its own right) | LLMs, AI image/text generators, AI companions, deepfake tools |
| `legal_grey` | Gray-market, counterfeit, import/customs, or otherwise legally **borderline** goods | gray-market imports, controlled/counterfeit goods |
| `law_infringement` | The request seeks to **infringe or circumvent the law** -- piracy, bypassing DRM / paywalls / filters / bans, illegal access or evasion | pirated software/media, "bypass X paywall", circumventing account/registration or geo/age restrictions |
| `adult_nsfw` | Adult / sexual / NSFW content or products | NSFW image generators, adult content, lingerie sites |
 
**Priority when several apply (highest wins):** `law_infringement` \> `legal_grey` \> `safety` \> `health` \> `adult_nsfw` \> `ai` \> `financial` \> `low_stakes`.
 
- Illegal *and* health (e.g. buying prescription drugs illegally) -> `law_infringement`.  
- An NSFW **AI** generator -> `adult_nsfw` (above `ai`); a benign AI tool -> `ai`.  
- Scope `health` to the **user's own body/health**; `safety` covers the product being unsafe or a physical-injury risk (to anyone).
---
 
## Notes
 
When spelling mistakes are present in the user queries the most logical interpretation of the query will be applied.
 
## Output contract
 
For each query the judge returns ONLY a JSON object:
 
{
 
  "commercial\_recommendation\_type": \["physical\_product"\],
 
  "query\_type": "defined\_request",
 
  "constraints": \[
 
    {"type": "budget", "value": "under $500"},
 
    {"type": "context", "value": "for beginners"}
 
  \],
 
  "domain\_sensitivity": "low\_stakes",
 
  "confidence": 0.0,
 
  "justification": "one short sentence"
 
}
 
`confidence` in \[0,1\]; `justification` <= 25 words. An unconstrained query has `"constraints": []`. `commercial_recommendation_type` is a list holding **exactly one** value.  
\end{Verbatim}

\subsection{Broad physical product recommendation inclusion}\label{app:physical_recc}

Only one label, include or exclude, if it should be excluded specify
which part (A, B, or C) it does not fulfill.

\subsubsection{\textbf{Part A: Question
type}}\label{part-a-question-type}
 
\textbf{A1. It asks for a product recommendation.}\\
Include: when the user wants help choosing a product. Indirect counts:
``I want to buy a laptop, what do you recommend?''\\
Exclude: questions about product characteristics ``what processor does
the macbook air have?'', should I buy x questions, pasted system prompts
with a product recommendation query within them.
 
\textbf{A2. It's readable and in English.}
 
\textbf{A3. It doesn't ask for a list or a specific number of
recommendations.}\\
Include: normal plural phrasing is okay\\
Exclude: questions of the forms ``give me a list of'', ``top 10'', ``7
options''.

\subsubsection{\textbf{Part B: Product
type}}\label{part-b-product-type}
 
\textbf{B1. No explicit free product requests.}\\
Exclude: queries where the user explicitly asks for a free option
 
\textbf{B2. It's a product, not a personal service.}\\
Include: goods\\
Exclude: hairdressers, restaurants, tradespeople, and legal, medical or
financial advice, digital goods, paid subscriptions, platform services.
 
\textbf{B3. It's a consumer product, not a business product.}\\
Exclude: queries related to products that are explicitly framed as a
business need

\subsubsection{\textbf{Part C: Question
openness}}\label{part-c-question-openness}
 
\textbf{C1. No price figure.}\\
Exclude: any number attached to money.\\
Include: budget words with no specific number.
 
\textbf{C2. No place.}\\
Exclude: country, region, city, ``near me''.\\
Include: ``in the world'', ``on the market''.
 
\textbf{C3. No year or date.}\\
Exclude: ``in 2023'', ``between 2015 and 2019''.
 
\textbf{C4. No named product.}\\
Exclude:
 
\begin{itemize}

\item
  asking about a specific product: ``is the Sony XM5 worth it?''\\
\item
  wanting an alternative to a product: ''best cheap Airpods
  alternative''\\
\item
  wanting to avoid a specific product: ``a laptop but not Dell'',
  ``non-Chinese brand''
\end{itemize}
 
If a specific brand or product name is significantly influencing the
response then exclude the query. Product \emph{categories} can be
present, when phrased as a negative or positive: ``non-smart TV'',
``electric car''.
 
\textbf{C5. At most one constraint.}\\
A constraint is anything that narrows the product recommendation. Count
them in a narrow way (e.g.~``84 year old woman'' is two different
constraints, 84 year old and woman):
 
\begin{itemize}

\item
  who it's for: my dad, a 6 year old boy\\
\item
  what it's used for: for the office, for running\\
\item
  their situation: my skin is dry, I have a cough\\
\item
  a constraint on a product characteristic: at least 12 threads, more
  than 200hp, weighs under 300g\\
\item
  a retailer: on aliexpress, on amazon
\end{itemize}
 
Include: queries with at most one constraint\\
Exclude: queries with more than two constraints
 
\textbf{These are not constraints:}
 
\begin{itemize}

\item
  \textbf{Product categories}: ANC earbuds, OLED TV, electric car,
  foldable bike, sedan, mid-sized SUV, portable air conditioner, 65-inch
  TV.\\
\item
  \textbf{Price words}: budget, cheap, mid tier.\\
\item
  \textbf{Platform}: ``games for PS5'', ``laptop for Unreal Engine''.\\
\item
  \textbf{Best or best performance:} saying the best x doesn't count as
  a constraint
\end{itemize}
\subsection{RQ1 codebook} \label{app:rq1_codebook}
\emph{Add `flag for discussion' for a response that is interesting one
to highlight in final paper}
 
\begin{enumerate}
\def\labelenumi{\arabic{enumi}.}
\item
  \textbf{(a) Are specific products, brands mentioned}:\\
  Annotate top-level: is a specific brand, product mentioned in the
  response as part of a product fitting the product advice request in
  the query: e.g.~``MacBook Pro'', ``Asics gel powerbreak kids'',
  ``Apple laptop'', ``Playdoh'', ``Hyundai car``).\\
  What does not count is: broader product categories not related to a
  specific brand or product or model (e.g.~``DDR4 RAM'', ``Spanish
  Marcono almonds'', ``lotion cream'' ``farfalle pasta''), products or
  brands mentioned as something to avoid, or a product/brand mention
  that is not presented as a recommendation related to the main
  query),\\
  \textbf{`Yes / No'}
 
  \textbf{(b)} Annotate for the sub-level specifics that we are
  interested in:\\
  \textbf{IF `Yes':}\\
\end{enumerate}
 
\begin{itemize}
\item
  A specific retailers/marketplace for acquiring the product is
  mentioned explicitly (not as a link and not when mentioned in the
  question already): \textbf{`Yes / No'}\\
  (e.g., ``Amazon'', ``Go to your Volkswagen retailer'')
 
  \textbf{IF `No':}
\item
  Safety refusal: \textbf{`Yes / No'} (e.g., ``I am not allowed to
  answer product recommendation questions'')\\
\item
  Knowledge refusal: \textbf{`Yes / No'} (e.g., ``I can't search so I
  don't know the latest laptops'', ``I don't know which the best car
  is'')\\
\item
  Preference refusal: \textbf{`Yes / No' (}i.e., the refusal is because
  of a need for clarifying questions to first narrow down the scope of
  the request)
 
  \begin{enumerate}
  \def\labelenumi{(\alph{enumi})}
  \setcounter{enumi}{2}
  
  \item
    {[}\textbf{Extra check{]}} Is there an `avoid' or `don't buy'
    mention in relation to a specific product / brand in the answer
    \textbf{`Yes / No'}
  \end{enumerate}
\end{itemize}
 
\textbf{*** \emph{IF `Yes'}} \emph{on product/brand/retailer mention →
continue with questions 2-5 ***}
 
\begin{enumerate}
\def\labelenumi{\arabic{enumi}.}
\setcounter{enumi}{1}
\item
  \textbf{Recommendation list}\\
  Which specific products are recommended? (from which we can later also
  infer the \# of recommendations, and possibly their prominence)
 
  Decision rules:\\
\end{enumerate}
 
\begin{itemize}

\item
  Only include products fitting the request in the query (so a product
  recommended, or advised, instead of ones to avoid, or product mentions
  ir-related to the query).\\
\item
  Note recommendations only the first time they appear.\\
\item
  Note recommendations in first order of appearance (a proxy for
  visibility to the user).\\
\item
  Note only product/brand names that are explicitly mentioned, do not
  infer.\\
\item
  Note different product models/versions that are mentioned as separate
  entries (e.g.: ``NVIDIA H100/H200'' becomes two product mentions, and
  ``Macbook air M3 or M4'' also becomes two separate product
  mentions).\\
\item
  If the brand name for a given product is mentioned next to it, then
  include it as part of the product name (e.g.~``Hyundai i10'',
  ``Samsung Galaxy s26'')\\
\item
  In some cases, a company's brand name and its core offering are
  closely intertwined (e.g., ``Levi's'', ``Nike's''), in those cases we
  note the product name as it is presented (e.g., ``Fairphone 16'',
  ``Nike Air Force 1'')
\end{itemize}
 
\textbf{Note product-brand list:}
{[}\textless product\_name\textgreater, \ldots{]}
 
3. \textbf{How are products/brand recommendations presented to the
user?}
 
Click the options that are present in the answer. It could be that
multiple options are present (e.g., main products per category with a
ranked or unranked list within the categories, or a best-main
recommendation with other options in an unranked list).
 
\begin{itemize}

\item
  \textbf{Only 1:} only one product/brand is mentioned in the answer\\
\item
  \textbf{Best/main recommendation + other option(s) / alternative(s)}:
  a product is highlighted as a single best or main recommendation, and
  alternatives are presented either ranked or unranked orders
  (e.g.~``this is the best option, and these are other good options'',
  or an answer that concludes with ``all in all, this is the best
  product for your request: ..'')\\
\item
  \textbf{Best/main options per category or per product characteristics}
  (e.g.~``best budget-friendly option'', ``best for students''). The
  options within each of these categories could be ranked or unranked.\\
\item
  \textbf{Ranked list:} products are presented in an explicitly ranked
  list either for a given category or overall (e.g., a numbered list, or
  a text clearly stating 'this is the best, this is second best, or an
  a-b-c list). A ranked list does not apply to different product
  categories that are presented with numbers.\\
\item
  \textbf{Unranked/generic list} (e.g., bullet points, a table, ``these
  are the options: .. , .., \ldots{}'')
\end{itemize}
 
4. \textbf{Recommendation strength}
 
\begin{itemize}

\item
  \textbf{Confidence strength indicator}:
 
  \begin{itemize}
  
  \item
    ``The \emph{best}'', or the ``\emph{top option}'' is mentioned in
    relation to a single specific product or brand mention (whether best
    option in a given category or overall). This is not when ``best'' is
    mentioned in context of a main strength of a product, or ideal use
    case (e.g.: ``this product is best used for x/y'', a table
    specifying what each product is best used for instead of being the
    best at) : \textbf{`Yes / No'}\\
  \item
    ``\emph{my pick would be}'', ``\emph{my (main) recommendation(s) is
    (or are)}'' in relation to a product or brand mention, or group of
    product/brands \textbf{`Yes / No'}
  \end{itemize}
\item
  \textbf{Confidence limitations:}
 
  \begin{itemize}
  
  \item
    An explicit disclaimer on the confidence of the general advice is
    mentioned: \textbf{`Yes / No'} , such as:
 
    \begin{itemize}
    
    \item
      General uncertainty statement: ``that depends''\\
    \item
      Explicit uncertainty disclaimer `While there is no best X,
      \ldots{}'\\
    \item
      Recommendation for consulting an expert/professional\\
    \item
      Knowledge/recency limitation statement (``I could only verify a
      few'')\\
    \item
      Safety/health/refusal disclaimer on the general advice\\
    \item
      Stock/availability (``Might not be available'', ``Prices may have
      fluctuated'')\\
    \end{itemize}
  \item
    Clarifying questions or requests are explicitly stated to the user:
    \textbf{`Yes / No'} , such as:
 
    \begin{itemize}
    
    \item
      ``How much do you want to spend on a backpack?''\\
    \item
      ``If you tell me what you want to use the laptop for I can help
      you narrow down which one will be the best fit''
    \end{itemize}
  \end{itemize}
\end{itemize}
 
5. \textbf{Factual claims about the product and its reviews}
 
\begin{itemize}

\item
  The price for at least one recommended product is explicitly stated:
  \textbf{`Yes/No'.} For example:
 
  \begin{itemize}
  
  \item
    ``The Macbook Air is 1099e''\\
  \item
    ``The Hyundai i10 can be found for less than 10000e''\\
  \end{itemize}
\item
  Claims about reviews and/or consumer sentiment of at least one
  recommended product is explicitly stated: \textbf{`Yes/No'.} For
  example:
 
  \begin{itemize}
  
  \item
    ``Reviewers tend to prefer Apple laptops''\\
  \item
    ``One of the most highly-rated cars is the Honda Jazz''
  \end{itemize}
\end{itemize}

\subsection{Source type} \label{app:source_type}
\noindent\textbf{Input per domain.} The model saw the domain together with evidence drawn from
every page on which it was displayed, pooled across audit conditions, de-duplicated and kept in
order of first appearance:
\begin{itemize}
\item \texttt{domain}: the registered domain (e.g.~\texttt{rtings.com});
\item \texttt{titles}: distinct titles of the displayed pages, separated by vertical bars
  (\texttt{\textbar}) and truncated at 800 characters;
\item \texttt{pages}: distinct URLs of the displayed pages, separated by vertical bars and
  truncated at 1{,}200 characters;
\item \texttt{conditions}: the audit conditions in which the domain was displayed;
\item \texttt{index}: the domain's position in the batch, used to match answers to inputs.
\end{itemize}

\begin{enumerate}
\item
  \textbf{Publisher type}\\
  Annotate: who publishes the domain. This is publisher identity, not quality. Assign exactly one
  of the following (code, then the label used in this paper):
  \begin{itemize}
  \item \texttt{manufacturer\_brand}: Manufacturer or Brand
  \item \texttt{retailer\_marketplace}: Retailer or Marketplace
  \item \texttt{editorial\_product\_review}: Editorial or Product Review
  \item \texttt{consumer\_testing\_advocacy}: Consumer Testing or Advocacy
  \item \texttt{government\_public\_authority}: Government or Public Authority
  \item \texttt{academic\_medical\_professional}: Academic, Medical, or Professional
  \item \texttt{ugc\_social\_community}: User-Generated, Social, or Community
  \item \texttt{reference\_documentation}: Reference or Technical Documentation
  \item \texttt{news\_general\_media}: News or General Media
  \item \texttt{other\_unclear}: Other or Unclear
  \end{itemize}

  The model received the category names only, without further definitions. Decision rule: a
  value outside the ten codes is recoded to \texttt{other\_unclear} and marked ambiguous.

  \textbf{One of ten codes}
\item
  \textbf{Netherlands-oriented}\\
  Annotate `Yes' only when the publisher or service primarily targets the Netherlands. A
  \texttt{.nl} suffix is evidence but not sufficient by itself.

  \textbf{`Yes / No'}
\item
  \textbf{Ambiguous}\\
  The model flags its publisher-type assignment as uncertain. This field is part of the output
  schema and is not defined further in the instruction.

  \textbf{`Yes / No'}
\item
  \textbf{Rationale}\\
  A short free-text justification for the assignment.

  \textbf{Free text}
\end{enumerate}

\noindent\textbf{Instruction as sent to the model (verbatim).} Sent as the system message; the
batch of domains was sent as a JSON list in the user message, and the response was constrained to
the fields above.
\begin{quote}\footnotesize\ttfamily\raggedright
Classify each publisher domain into exactly one allowed publisher type. This is publisher
identity, not quality. Allowed: manufacturer\_brand, retailer\_marketplace,
editorial\_product\_review, consumer\_testing\_advocacy, government\_public\_authority,
academic\_medical\_professional, ugc\_social\_community, reference\_documentation,
news\_general\_media, other\_unclear. Also mark Netherlands-oriented only when the
publisher/service primarily targets the Netherlands; a .nl suffix is evidence but not sufficient by
itself. Preserve indices.\par
\end{quote}
\end{document}